\documentclass[aps,pra,twocolumn,showpacs,superscriptaddress,groupedaddress]{revtex4}

\usepackage{graphicx}
\usepackage{amsthm}
\usepackage[toc,page,title,titletoc,header]{appendix}
\usepackage{graphicx}
\usepackage{epstopdf}
\usepackage{CJK}
\usepackage{mathpazo}

\theoremstyle{plain}

\theoremstyle{definition}

\theoremstyle{remark}

\newcommand{\PRA}[3] {Phys. Rev. A {\bf #1}, #2 (#3)}

\newcommand{\PRE}[3] {Phys. Rev. E {\bf #1}, #2 (#3)}
\newcommand{\PRL}[3] {Phys. Rev. Lett. {\bf #1}, #2 (#3)}
\newcommand{\JPA}[3] {J. Phys. A {\bf #1}, #2 (#3)}
\newcommand{\JPB}[3] {J. Phys. B {\bf #1}, #2 (#3)}

\newcommand{\EPL}[3] {Europhysics Letters {\bf #1}, #2 (#3)}
\newcommand{\EPJD}[3] {Eur. Phys. J. D {\bf #1}, #2 (#3)}
\newcommand{\EPJB}[3] {Eur. Phys. J. B {\bf #1}, #2 (#3)}
\newcommand{\EPJP}[3] {Eur. Phys. J. Plus {\bf #1}, #2 (#3)}

\newcommand{\PS}[3] {Phys. Scr. {\bf #1} #2 (#3)}

\newcommand{\NJP}[3] {New J. Phys. {\bf #1}, #2 (#3)}

\newcommand{\Science}[3] {Science {\bf #1}, #2 (#3)}
\newcommand{\JAP}[3] {J. Appl. Phys. {\bf #1}, #2 (#3)}

\newcommand{\PNAS}[3] {PNAS {\bf #1}, #2 (#3)}
\newcommand{\NP}[3] {Nat. Phys. {\bf #1}, #2 (#3)}

\newcommand{\PA}[3] {Physica A {\bf #1}, #2 (#3)}

\newcommand{\IJMPB}[3] {Int. J. Mod. Phys. B {\bf #1}, #2 (#3)}
\newcommand{\AnnP}[3] {Ann. Phys. {\bf #1}, #2 (#3)}
\newcommand{\Entropy}[3] {Entropy {\bf #1}, #2 (#3)}
\newcommand{\Scirep}[3] {Sci. Rep. {\bf #1}, #2 (#3)}
\newcommand{\QIP}[3] {Quantum Inf. Process {\bf #1}, #2 (#3)}

\begin{document}

\begin{CJK*}{GB}{gbsn}

\title{Multilevel quantum Otto heat engines with  identical particles}

\author{X. L. Huang}
\email{huangxiaoli1982@foxmail.com}
\author{D. Y. Guo}
\affiliation{School of physics and electronic technology,
Liaoning Normal University, Dalian, 116029, China}
\author{S. L. Wu}
\affiliation{School of Physics and Materials Engineering, Dalian Nationalities University, Dalian 116600 China}
\author{X. X. Yi}
\affiliation{Center for Quantum Sciences and School of Physics, Northeast Normal University, Changchun 130024, China}

\date{\today}

\begin{abstract}
A quantum Otto heat engine is studied with  multilevel identical particles trapped in one-dimensional box potential as  working substance. The symmetrical wave function for Bosons and the anti-symmetrical wave function for Fermions are considered. In two-particle case, we focus on the ratios of $W^i$ ($i=B,F$) to $W_s$, where $W^B$ and $W^F$ are the work done by two Bosons and Fermions respectively, and $W_s$ is the work output of a single particle under the same conditions. Due to the symmetric of the wave functions, the ratios are not equal to $2$. Three different regimes, low temperature regime, high temperature regime, and intermediate temperature regime, are analyzed, and the effects of energy level number and the differences between the two baths are calculated.  In the multiparticle case, we calculate the ratios of $W^i_M/M$ to $W_s$, where $W^i_M/M$ can be seen as the average work done by a single particle in multiparticle heat engine.
 For other working substances whose energy spectrum have the form of $E_n\sim n^2$, the results are similar. For the case $E_n\sim n$, two different conclusions are obtained.
\end{abstract}

 \maketitle

\end{CJK*}

\section{Introduction}

In recent years, with the development of quantum information theory \cite{Nielsenbook} and the control technics on a single atom, more and more attentions have been paid to the study of quantum heat engines, which use quantum systems as working substances and are the quantum mechanics generalizations of classical heat engines.
In the classical thermodynamics, heat engines are  devices which can convert heat to work or transfer the energy from one place to another, and they are composed of several thermodynamics processes. There are four basic processes in classical thermodynamics, i.e., adiabatic process with fixed entropy, isothermal process with fixed temperature, isochoric process with fixed volume, and isobaric process with fixed pressure. The adiabatic process is the first process generalized to quantum case by quantum adiabatic theorem \cite{Sakuraibook}, and it is more rigorous than classical adiabatic process. The classical or quantum isothermal process is similar and the working substance is always thermal equilibrium with a heat bath with fixed temperature. The generalizations of the other two processes meet some difficulties because it usually  does not have the concepts of volume and pressure in many quantum models for example, spin and harmonic oscillator. This difficulty is overcome by the concepts of the generalized coordinate and generalized force \cite{Kieu2004PRL,Kieu2006EPJD,Quan2007PRE,Quan2009PRE}. So far, the four basic processes have been all generalized to quantum mechanics regimes. And all kinds of quantum systems, such as a single two-level atom \cite{Wang2011PRE,Wang2012PRE,Niu2015IJMPB,Huang2017QIP,Leggio2016PRE,Munoz2012PRE,Correa2016Entropy,Yuan2014PRE}, multilevel atom \cite{Quan2005PRE,Gaveau2010PRE,Pena2016PRE,Cakmak2016EPJD}, harmonic oscillator \cite{Wang2015PRE,Insinga2016PRE,Abah2016EPL}, non-Hermit system with $\mathcal {PT}$ symmetric \cite{Lin2016JPA,Gardas2016Scirep}, and coupled system \cite{Zhang2007PRA,Thomas2011PRE,Huang2013PRE,Thomas2014EPJB,Altintas2014PRE,Cakmak2016EPJP,Altintas2015PRE,Azimi2014NJP,Huang2018PA,Basu2017PRE,Chotorlishvili2016PRE,Ivanchenko2015PRE,Zhao2017QIP,Huan2013PS}, can be used as the working substances. Many quantum heat engine models may be or have been realized in experiments \cite{Abah2012PRL,Blickle2012NP,Fialko2012PRL,Zhang2014PRL,Zhang2015PRE,Ian2014JPB,Robnagel2016Science,Alickia2017AP}.

Among these studies, one motivation is to study the effects of quantum mechanics resource on the thermodynamical quantities. For example, non-equilibrium heat reservoirs can improve the heat engine efficiencies and the work extractions \cite{Scully2003Science,Huang2012PRE,Robnagel2014PRL,Manzano2016PRE,Zhang2014JPA,Long2015PRE,Mehta2012AP,Scully2011PNAS}. In the case of coupled quantum systems as the working substances, the second law of thermodynamics can not be violated although the entanglements, a full non-classical resource in many body systems, exist in the systems. Moreover, the heat current can have the different directions between the total system and the local subsystems \cite{Thomas2011PRE,Huang2013PRE,Huang2014EPJP}.

The concept of indentity is another basic concept in quantum mechanics. It makes the ideal quantum gases at low temperature have different properties, i.e., the Bose-Einstein condensation and the Fermi sphere \cite{Pathriabook}. The effects of identical particles as  working substances are first noticed in Refs. \cite{RWang2012PRE,Wang2011JAP}, in which multilevel $N$-particle Fermions are used as  working substances to complete an isoenergetic cycle. The Pauli exclusion principle is considered in the cycle. In this paper, we consider ideal identical multilevel particles as  working substances to construct  quantum Otto cycles. We consider the symmetrical wave function for Bosons and the anti-symmetrical wave function for Fermions, and discuss the effects of different particle statistics and the energy level numbers on the thermodynamical quantities of the cycle.

This paper is organized as follows. In Sec. \ref{sec:WS}, the working substance and the cycle are introduced. The main results of the two-particle case and analyses are given in Sec. \ref{sec:Result} and the generalizations to the multiparticle case are discussed in Sec. \ref{sec:MC}. The discussions about other working substances and the conclusions are presented in Sec. \ref{sec:Con}.

\section{Two-particle case} \label{sec:TC}
\subsection{Working substance and the cycle}\label{sec:WS}

\begin{figure}
\includegraphics*[width=0.6\columnwidth,bb=180 320 440 550]{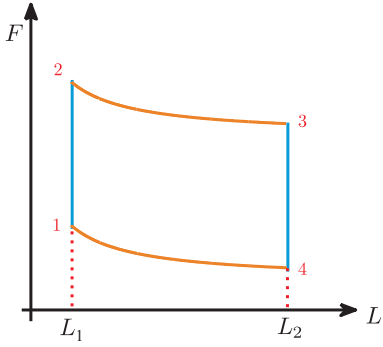}
\caption{Description of a quantum Otto cycle, where $F=\sum_np_n\frac{\partial E_n}{\partial L}$ is the generalized force, the generalization of the pressure in classical thermodynamics, of the working substance.  }\label{fig:10}
\end{figure}

We first consider two noninteracting identical particles trapped in one-dimensional (1D) box potential as the working substance. For a single particle, the eigenenergies read
\begin{eqnarray}
E_n=\frac{n^2\pi^{2}\hbar^{2}}{2mL^{2}}, \label{eq:10}
\end{eqnarray}
with the corresponding single particle wave functions $\varphi_{n}(x)$, where $L$ is the potential width (the generalized coordinate) and $m$ is the mass of the particle. Note that $E_n$ and $\varphi_n$ are the single particle eigenenergies and wave function respectively. The working substances of our cycle in this section are two identical particles.

A quantum Otto cycle consists of two quantum adiabatic processes and two quantum isochoric processes (see Fig. \ref{fig:10}). Starting from point $1$, $1$ to $2$ is a quantum isochoric process. At point $1$, the potential width is $L_{1}$, and the energy levels of the single particle are $E_{n}(L_{1})$. The occupation probability of each energy level for the two-identical-particle system (the working substance) is $p_{n}^{c}$. During this isochoric process, the working substance keeps contact with a hot heat bath at $T_{h}$. At point $2$, the working substance reaches the thermal equilibrium with the bath. The occupation probability of each energy level for the two-identical-particle system becomes $p_{n}^{h}$. The internal energy of the working substance can be obtained as
\begin{eqnarray}
U_{2}=-\frac{\partial}{\partial\beta}\ln{\mathcal Z_{h}}, \label{eq:20}
\end{eqnarray}
where $\mathcal Z_{h}=\mathcal Z_{h}(T_{h},L_{1})$ is the partition function at point $2$ and $\beta$ is the inverse temperature, i.e., $\beta=1/kT$. $2\rightarrow3$ is a quantum adiabatic process where the potential width changes from $L_{1}$ to $L_{2}$. This process is so slow  that each occupation probability does not change. According to the energy spectrum of the single particle given in Eq. (\ref{eq:10}), we obtain the internal energy at point $3$ as
\begin{eqnarray}
U_{3}=\left(\frac{L_{1}}{L_{2}}\right)^{2}U_{2}. \label{eq:30}
\end{eqnarray}
$3\rightarrow4$ is another quantum isochoric process, in which the working substance is coupled with a cold bath at $T_{c}$. In enough  time, it reaches thermal equilibrium with the cold bath and then the internal energy becomes
\begin{eqnarray}
U_{4}=-\frac{\partial}{\partial\beta}\ln{\mathcal Z_{c}}, \label{eq:40}
\end{eqnarray}
where $\mathcal Z_{c}=\mathcal Z_{c}(T_{c},L_{2})$ is the partition function at point $4$. The last stage $4\rightarrow1$ is a quantum adiabatic process in which the potential width changes from $L_{2}$ back to $L_{1}$ and each occupation probability of the two-particle system $p_{n}^{c}$ is maintained. Similar to the second stage, the internal energy at point $1$ can be arranged as
\begin{eqnarray}
U_{1}=\left(\frac{L_{2}}{L_{1}}\right)^{2}U_{4}. \label{eq:50}
\end{eqnarray}

Based on the above description, some heat $Q_{h}$ are absorbed from the hot bath at stage $1$, and some of them, i.e., $Q_{c}$, are released to the cold bath at stage $3$. The heat transferred can be calculated according to the change in internal energy as
\begin{eqnarray}
&&Q_{h}=U_{2}-U_{1}=U_{2}-\left(\frac{L_{2}}{L_{1}}\right)^{2}U_{4}, \label{eq:60}\\
&&Q_{c}=U_{3}-U_{4}=\left(\frac{L_{1}}{L_{2}}\right)^{2}U_{2}-U_{4}. \label{eq:70}
\end{eqnarray}
After a whole cycle the work done by the heat engine can be obtained according to the first law of thermodynamics as
\begin{eqnarray}
W=Q_{h}-Q_{c}. \label{eq:80}
\end{eqnarray}
At last the heat engine efficiency is
\begin{eqnarray}
\eta=\frac{W}{Q_{h}}=1-\frac{Q_{c}}{Q_{h}}=1-\left(\frac{L_{1}}{L_{2}}\right)^{2}. \label{eq:90}
\end{eqnarray}
We can see that the efficiencies are the same for both Bosons and Fermions and they are the same as the result of a single particle as the working substance. This efficiency is always less than the Carnot efficiency. As a result, we don't consider the efficiency in the following discussions.
Moreover, the ratio of $L_2$ to $L_1$ can be seen as the adiabatic compression ratio, i.e.,
\begin{eqnarray}
R=\frac{L_2}{L_1},
\end{eqnarray}
is another important parameter in the cycle.

\subsection{Results and analyses}\label{sec:Result}
\begin{figure}
\includegraphics*[width=0.95\columnwidth,bb=65 230 505 565]{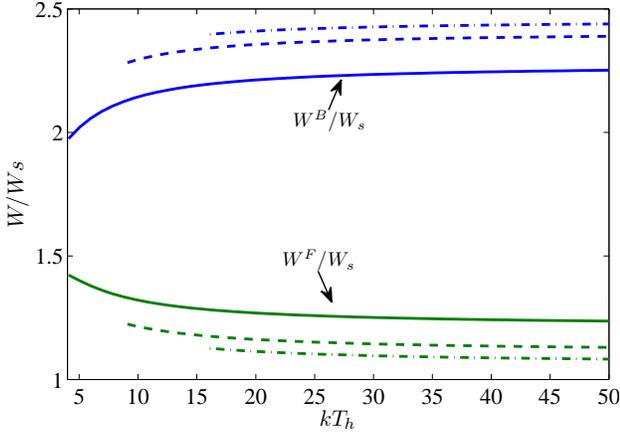}
\caption{(Color online) The ratios of $W^i/W_s$ as a function of $kT_h$ for different $R$. The blue lines denote the results of Bosons while the green lines are the ones of Fermions. The adiabatic expansion ratios $R$ are $2$, $3$, and $4$, corresponding to solid line, dash line and dot-dash line.  Other parameters are chosen as ${\hbar^2\pi^2}/{2mL_1^2kT_c}=1$ (the intermediate temperature regime, see the discussions below) and $kT_h$ is plotted in units of $kT_c$.     }\label{fig:20}
\end{figure}

We first consider the three-level case, i.e., only the lowest three levels of single-particle states are involved. The wave functions of the three single-particle states are denoted by $\varphi_{1}(x)$,  $\varphi_{2}(x)$ and $\varphi_{3}(x)$ with the corresponding eigenvalues $E_{1}(L)=\frac{\hbar^{2}\pi^2}{mL^{2}}$, $E_{2}(L)=\frac{4\hbar^{2}\pi^2}{mL^{2}}$ and $E_{3}(L)=\frac{9\hbar^{2}\pi^2}{mL^{2}}$. In the two identical particles cases, the total wave functions should be symmetrical for Bosons and anti-symmetrical for Fermions. For the Bosons, the symmetrical wave functions and their corresponding eigenergies are
\begin{eqnarray}
    \psi_{1}^{B}{=}\varphi_{1}(x_{1})\varphi_{1}(x_{2}),~~~~~~~~~~~~~~~~~~~~~~~~~~  E_{1}^{B}{=}\frac{2\hbar^{2}\pi^{2}}{2mL^{2}}, \nonumber\\
    \psi_{2}^{B}{=}\frac{1}{\sqrt2}[\varphi_{1}(x_{1})\varphi_{2}(x_{2}){+}\varphi_{2}(x_{1})\varphi_{1}(x_{2})], E_{2}^{B}{=}\frac{5\hbar^{2}\pi^{2}}{2mL^{2}}, \nonumber\\
    \psi_{3}^{B}{=}\varphi_{2}(x_{1})\varphi_{2}(x_{2}),~~~~~~~~~~~~~~~~~~~~~~~~~~  E_{3}^{B}{=}\frac{8\hbar^{2}\pi^{2}}{2mL^{2}}, \nonumber\\
    \psi_{4}^{B}{=}\frac{1}{\sqrt2}[\varphi_{1}(x_{1})\varphi_{3}(x_{2}){+}\varphi_{3}(x_{1})\varphi_{1}(x_{2})], E_{4}^{B}{=}\frac{10\hbar^{2}\pi^{2}}{2mL^{2}}, \nonumber \\
    \psi_{5}^{B}{=}\frac1{\sqrt2}[\varphi_{2}(x_1)\varphi_{3}(x_{2}){+}\varphi_{3}(x_{1})\varphi_{2}(x_{2})], E_{5}^{B}{=}\frac{13\hbar^{2}\pi^{2}}{2mL^{2}}, \nonumber\\
    \psi_{6}^{B}{=}\varphi_{3}(x_{1})\varphi_{3}(x_{2}),~~~~~~~~~~~~~~~~~~~~~~~~~~  E_{6}^{B}{=}\frac{18\hbar^{2}\pi^{2}}{2mL^{2}}, 
\label{eq:100}
\end{eqnarray}
where $x_{1}$ and $x_{2}$ are the coordinates of the first and second particle respectively. As a result, the partition function reads
\begin{eqnarray}
\mathcal Z_{i}^{B}(T,L)=\sum_{n=1}^6\exp[-\beta_{i}E_{n}^{B}(L)], \label{eq:110}
\end{eqnarray}
where $i=c$ and $h$. Similarly, for the Fermions, the anti-symmetrical wave functions and the corresponding eigenenergies are
\begin{eqnarray}
\psi_{1}^{F}{=}\frac{1}{\sqrt2}[\varphi_{1}(x_{1})\varphi_{2}(x_{2}){-}\varphi_{2}(x_{1})\varphi_{1}(x_{2})],
E_{1}^{F}{=}\frac{5\hbar^{2}\pi^{2}}{2mL^{2}}, \nonumber\\
\psi_{2}^{F}{=}\frac{1}{\sqrt2}[\varphi_{1}(x_{1})\varphi_{3}(x_{2}){-}\varphi_{3}(x_{1})\varphi_{1}(x_{2})],
E_{2}^{F}{=}\frac{10\hbar^{2}\pi^{2}}{2mL^{2}}, \nonumber\\
\psi_{2}^{F}{=}\frac{1}{\sqrt2}[\varphi_{2}(x_{1})\varphi_{3}(x_{2}){-}\varphi_{3}(x_{1})\varphi_{2}(x_{2})],
E_{3}^{F}{=}\frac{13\hbar^{2}\pi^{2}}{2mL^{2}}. \label{eq:120}
\end{eqnarray}
The partition function is
\begin{eqnarray}
\mathcal Z_{i}^{F}(T,L)=\sum_{n=1}^3\exp[-\beta_{i}E_{n}^{F}(L)]. \label{eq:130}
\end{eqnarray}
Putting Eqs. (\ref{eq:110}) and (\ref{eq:130}) into Eqs. (\ref{eq:20}) and (\ref{eq:40}), we can calculate the work done by the two three-level identical particles, i.e., $W^{B}$ for Bosons and $W^{F}$ for Fermions. In this section we focus on the ratio of $W^{i}(i=B,F)$ to $W_{s}$, where $W_{s}$ is the work done by a single particle as the working substance under the same conditions and it can be obtained by using single particle partition function in Eqs. (\ref{eq:20}) and (\ref{eq:40}). Here the same conditions mean the same external conditions, i.e., $L_1$, $L_2$, $T_c$ and $T_h$ are same for a single particle and two identical particles as working substances. The analytical expressions of the ratios are complex so we only give the numerical results here. They are shown in Fig. $2$ with different adiabatic compression ratios $R$. We note that the positive work condition of the quantum Otto heat engine is $T_{h}>R^{2}T_{c}$ for a single particle trapped in 1D box potential as the working substance. This is also true for two identical three-level particles as the working substance. Moreover, we can see that in vast majority areas $W^{B}/W_{s}$ is larger than 2,  however, $W^{F}/W_{s}$ is smaller than $2$. The larger adiabatic compression ratio $R$ makes this phenomenon more obviously. In other words, in most cases the work done by the two identical three-level particles under Bose statistics is larger than the two three-level particles under the Maxwell-Boltzmann statistics. But the work done by the particles under Fermi statistics is smaller. When $R$ and $kT_{h}$ are large enough, $W^{F}/W_{s}$ approaches to $1$, which means that the work done by the two Fermions is similar to a single particle in three-level case.

In the following discussions, the effects of adiabatic compression ratios $R$ are similar, i.e., larger $R$ makes $W^B/W_s$ larger for Bosons and $W^F/W_s$ smaller for Fermions, and the positive work condition is determined by $T_h>R^2T_c$. As a result, we only consider a simple case $R=2$ in the following discussions.

\begin{figure}
\includegraphics*[width=0.95\columnwidth,bb=90 260 480 562]{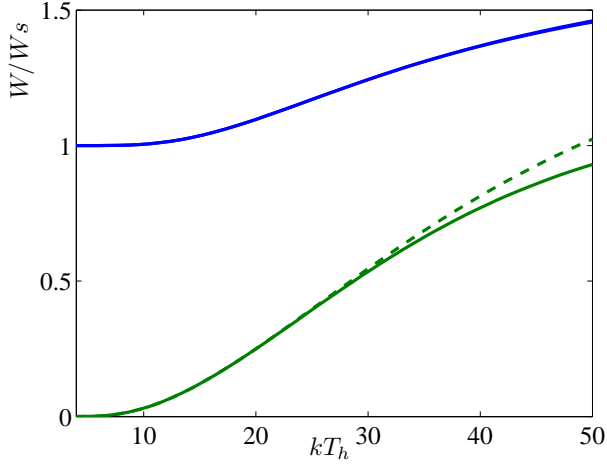}
\caption{(Color online) The ratios of $W^i/W_s$ as a function of $kT_h$ in the low temperature regime. The blue lines denote the results of Bosons and the green lines are the ones of Fermions. The solid line is $N=3$ and dash line is  $N=4$. Other parameters are ${\hbar^2\pi^2}/{2mL_1^2kT_c}=20$, and $R=2$. $kT_h$ is plotted in units of $kT_c$. }\label{fig:30}
\end{figure}

Next we consider $N$-level cases, where $N>3$. It is more complex than three-level case and we only write down the partition functions here. For two $N$-level identical particles, symmetrical wave functions lead to the following partition function as,
\begin{eqnarray}
\mathcal Z_{i}^{B}(T,L)=\sum_{i=1}^N\sum_{j=i}^N\exp[-\frac{\hbar^{2}\pi^{2}}{2mkTL^{2}}(i^{2}+j^{2})], \label{eq:140}
\end{eqnarray}
and anti-symmetrical wave functions lead to
\begin{eqnarray}
\mathcal Z_{i}^{F}(T,L)=\sum_{i=1}^{N-1}\sum_{j=i+1}^N\exp[-\frac{\hbar^{2}\pi^{2}}{2mkTL^{2}}(i^{2}+j^{2})]. \label{eq:150}
\end{eqnarray}
It can be easily checked that when $N=3$ , Eqs. (\ref{eq:140}) and (\ref{eq:150}) reduce to Eqs. (\ref{eq:110}) and (\ref{eq:130}) . Following the similar procedure, we can obtain $W^{i}/W_{s}$ under different conditions. Here we consider three different cases: low temperature regime $\hbar^{2}\pi^{2}/2mL_{1}^{2}kT_{c}\gg1$, high temperature regime $\hbar^{2}\pi^{2}/2mL_{1}^{2}kT_{c}\ll1$ and intermediate temperature regime $\hbar^{2}\pi^{2}/2mL_{1}^{2}kT_{c}\sim1$. We calculate the ratio of $W^{i}$ to $W_{s}$ in these three regimes and study the effects of energy level number $N$ and particle statistics on the work extraction. Fig. \ref{fig:30} shows the results of low temperature limit. In this figure we find that when $kT_{h}$ is low (sightly higher than the positive work condition $R^{2}kT_{c}$), $W^{i}/W_{s} (i=B$ and $F)$ is $1$ for Bosons and $0$ for Fermions. This can be understood as follows. When the temperature is low enough, only the lowest two energy levels are occupied. In other words, only the lowest two energy levels are effective for the work extraction. For Bosons, the lowest two energy levels are $E_{1}^{B}=\frac{2\hbar^{2}\pi^{2}}{2mL^{2}}$ and $E_{2}^{B}=\frac{5\hbar^{2}\pi^{2}}{2mL^{2}}$. The energy level difference is $\Delta E_{12}^{B}=\frac{3\hbar^{2}\pi^{2}}{2mL^{2}}$, which is same as the one of a single particle. However, the ones for Fermions are $\Delta E_{12}^{F}=E_{2}^{F}-E_{1}^{F}=\frac{5\hbar^{2}\pi^{2}}{2mL^{2}}$, which is larger than the single particle. In the low temperature case, this makes it much harder to occupy the first excited state. As a result, the work done can be ignored for Fermions. When $kT_{h}$ increases, the ratios $W^{i}/W_{s}$ increase for both Bosons and Fermions. But the effects of energy level number $N$ is weak. Here we only give the results of $N=3$ and $N=4$. For Bosons these two lines are coincident. The results of $N\geq5$ are nearly same to the ones of $N=4$.
\begin{figure}
\includegraphics*[width=0.49\columnwidth,bb=85 260 480 562]{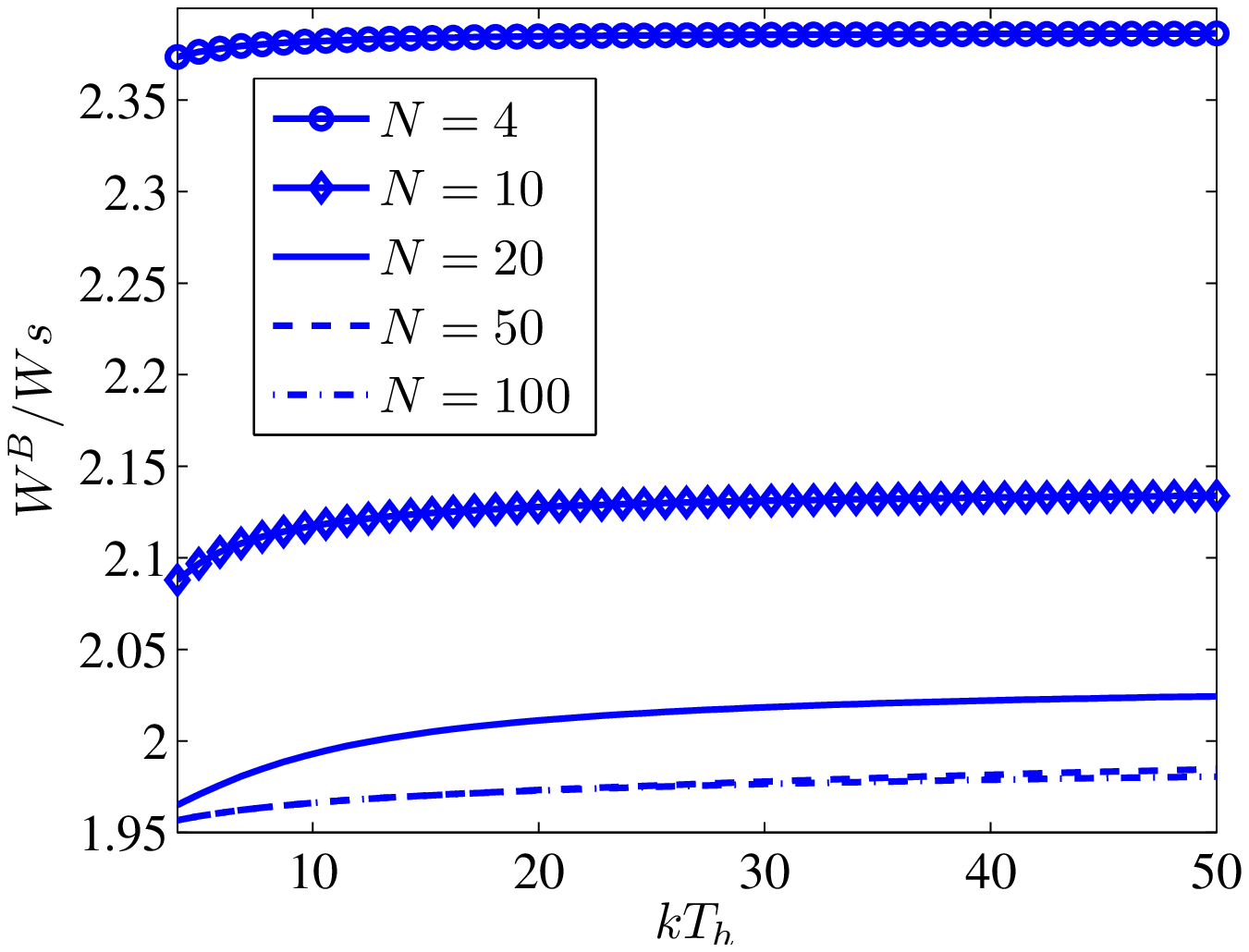}
\includegraphics*[width=0.49\columnwidth,bb=85 260 480 562]{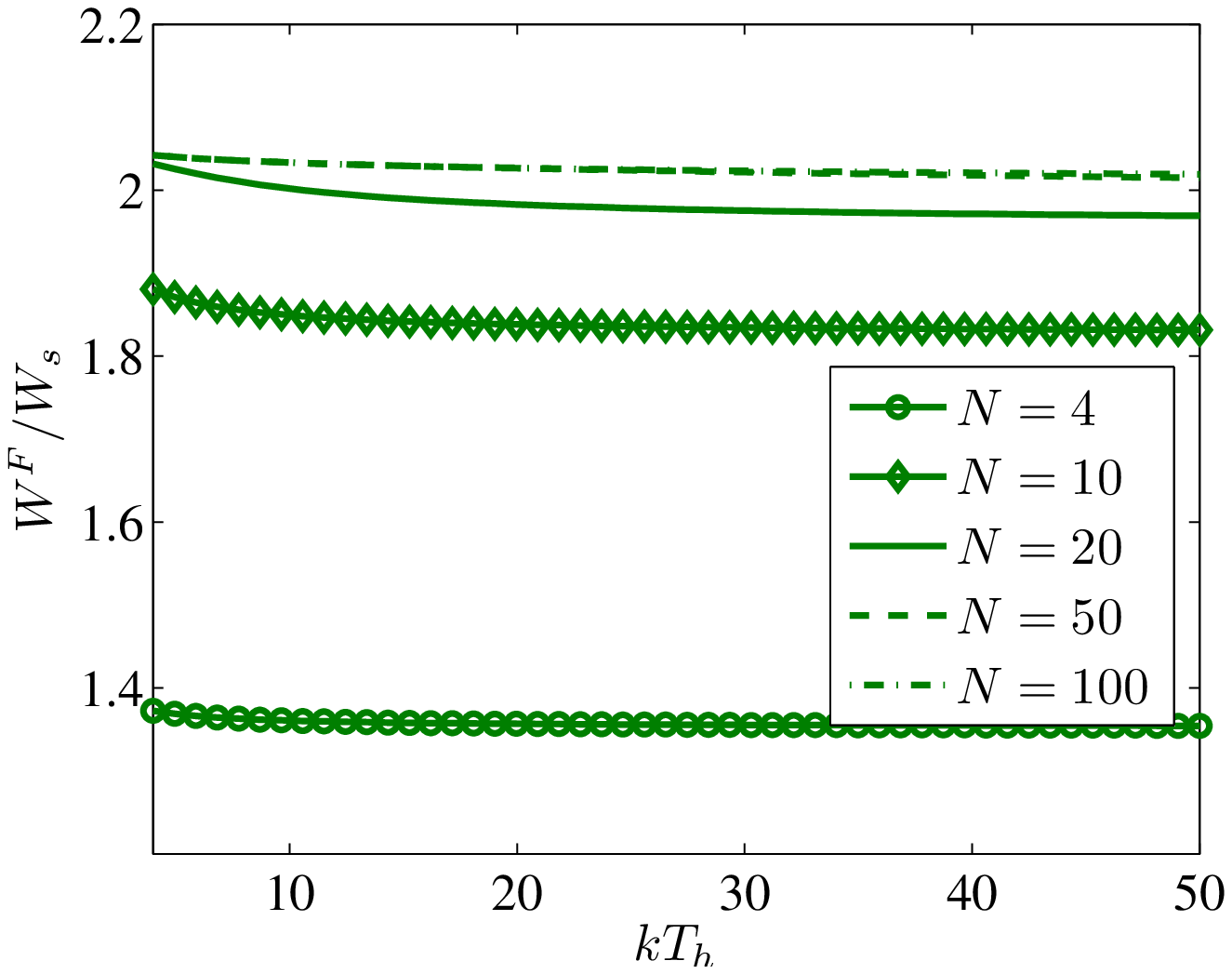}
\caption{The ratios of $W/W_s$ as a function of $kT_h$ for Bosons (left figure) and Fermions (right figure) with different $N$ in the high temperature regime. Other parameters are ${\hbar^2\pi^2}/{2mL_1^2kT_c}=0.05$, $R=2$ and $kT_h$ is plotted in units of $kT_c$. }\label{fig:40}
\end{figure}

The results in high temperature limit are quite different, which are shown in Fig. \ref{fig:40}. We first observe that the results of this case are energy level number $N$ sensitive. When $N$ is small, $W^{B}/W_{s}$ is larger than $2$ but $W^{F}/W_{s}$ is smaller than $2$. When $N$ increases, the ratio decreases for Bosons and increases for Fermions. When $N$ is large enough, this ratio can be slightly smaller than $2$ for Bosons but slightly larger than $2$ for Fermions. The effects of $kT_{h}$ on this ratio with fixed $N$ are weak, which means that we have nearly the same ratios for different $kT_{h}$ with fixed $N$. When $N>100$ the results do not change obviously with our parameters, which indicate that it can be seen as the result of $N\rightarrow\infty$. In this case, both for Bosons and Fermions, $W^{i}/W_{s}$ approaches $2$, which can be understood as that the quantum statistics tends to become the classical statistics in the high temperature limit for particles trapped in $1$D box potential.
\begin{figure}
\includegraphics*[width=0.49\columnwidth,bb=85 260 480 562]{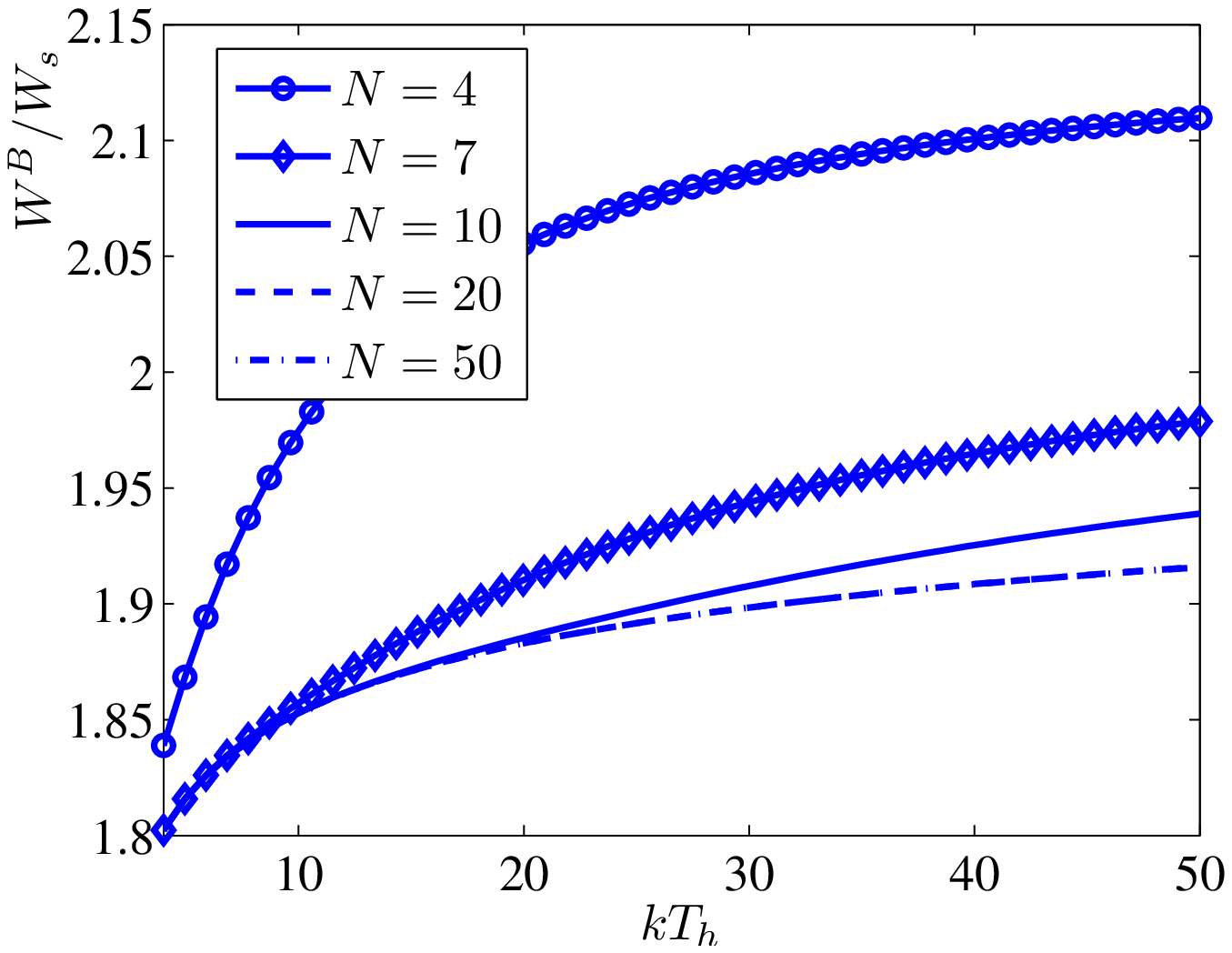}
\includegraphics*[width=0.49\columnwidth,bb=85 260 480 562]{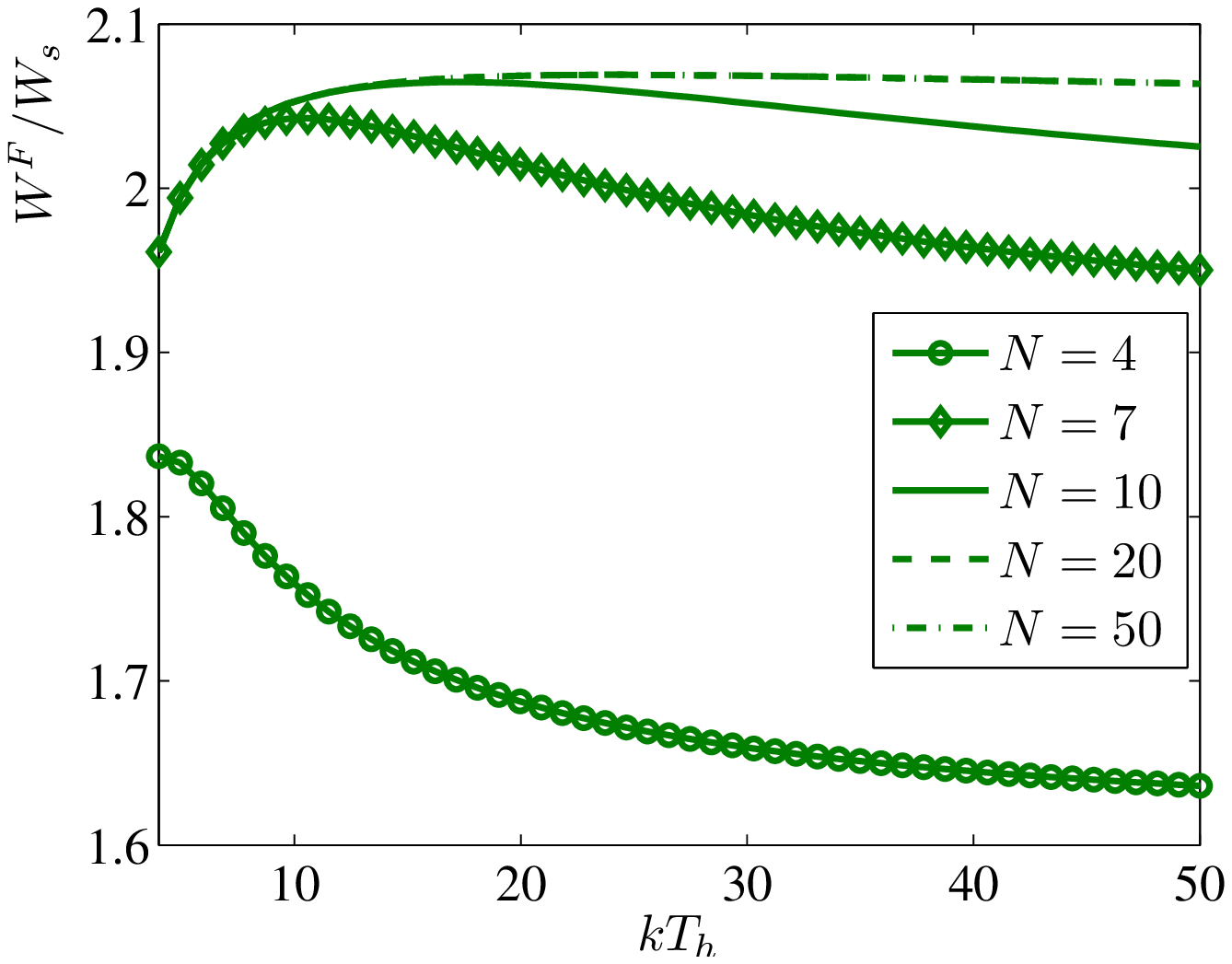}
\caption{The ratios of $W/W_s$ as a function of $kT_h$ for Bosons (left figure) and Fermions (right figure) with different $N$ in the intermediate temperature regime. Other parameters are ${\hbar^2\pi^2}/{2mL_1^2kT_c}=1$, $R=2$ and $kT_h$ is plotted in units of $kT_c$. }\label{fig:50}
\end{figure}

In the intermediate temperature regime, the results are more complex, which are shown in Fig. \ref{fig:50}. Similar to the high temperature case, the results are energy level number $N$ sensitive too. For Bosons as  working substances, $W^{i}/W_{s}$ is near $2$ when $N=4$. Then larger $N$ makes this ratio smaller, and for the large enough $N$, this ratio is smaller than $2$. $kT_{h}$ can make the ratios increase in all cases. For Fermions as working substances, $W^{i}/W_{s}$ is smaller than $2$ when $N=4$, and the larger $N$ becomes, the larger this ratio reaches. When $N$ is large enough, $W^{i}/W_{s}$ is near $2$. The effects of $kT_{h}$ are different. It may make this ratio decrease (for example when $N=4$) or has a nonmonotonic behavior (first increase and then decrease). When $N>50$, the results can be seen as $N\rightarrow\infty$.

Based on the results of high and intermediate temperatures cases, we confirm the conditions that quantum statistics trends to the classical statistics are both high temperature and infinitely many energy levels. In the finite energy levels case, for example, $N=4$ in Fig. \ref{fig:40}, $W^i/W_s$ is far from 2 in the high temperature limit. In the intermediate regime, when $T_h$ is large enough, $W_i/W_s$ are slightly different from 2 due to the intermediate temperature $T_c$ in the cycle.

\section{Multiparticle case} \label{sec:MC}
\begin{figure}
\includegraphics*[width=0.6\columnwidth,bb=70 260 500 560]{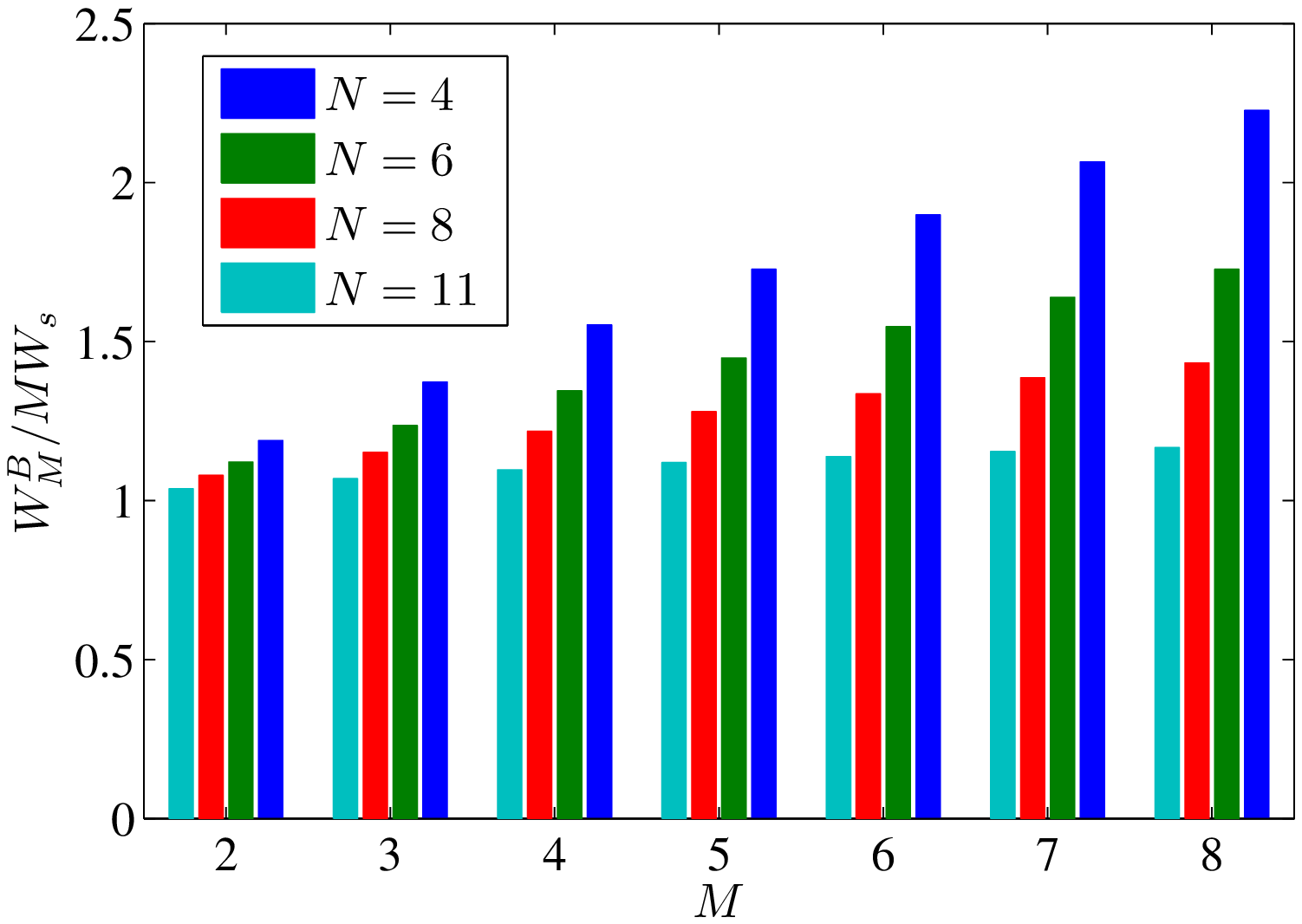}
\includegraphics*[width=0.6\columnwidth,bb=70 260 500 560]{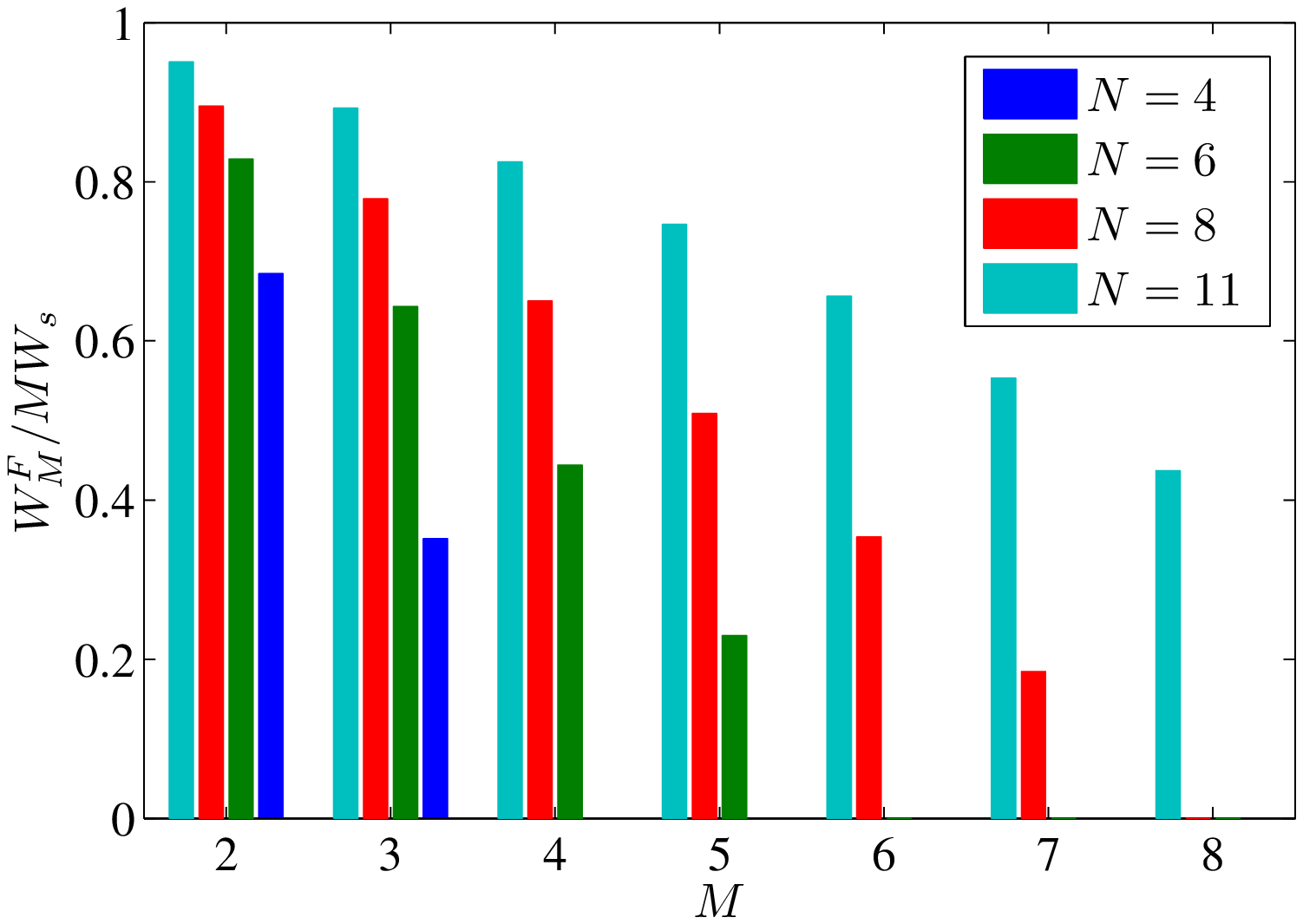}
\caption{(Color online) The ratios of $W^i_M/M$ to $W_s$ as a function of particle number $M$ for different energy level numbers $N$ in high temperature regime. Other parameters are ${\hbar^2\pi^2}/{2mL_1^2kT_c}=0.05$, $R=2$ and $kT_h=5kT_c$ }\label{fig:60}
\end{figure}
\begin{figure}
\includegraphics*[width=0.6\columnwidth,bb=70 260 500 560]{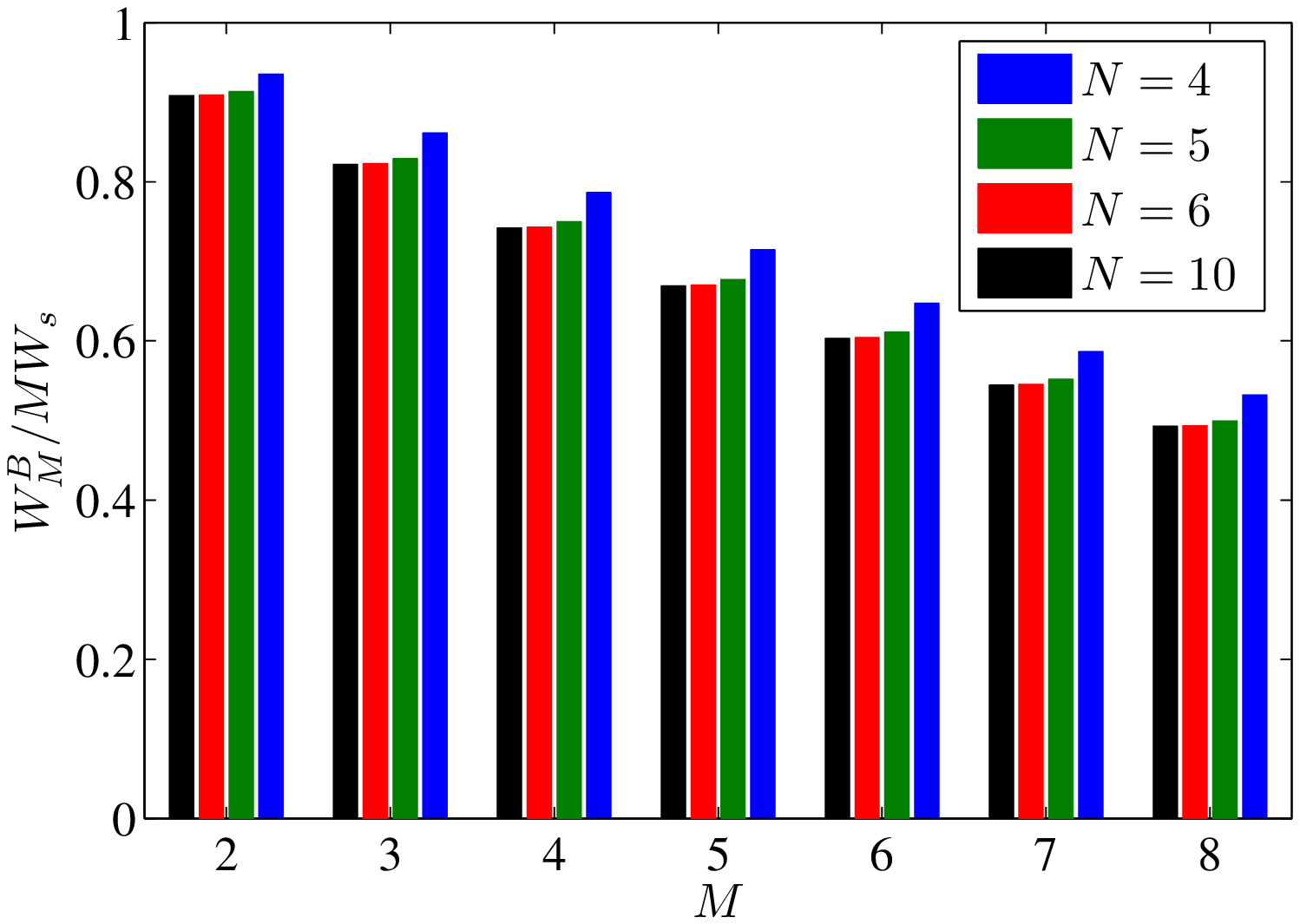}
\includegraphics*[width=0.6\columnwidth,bb=70 260 500 560]{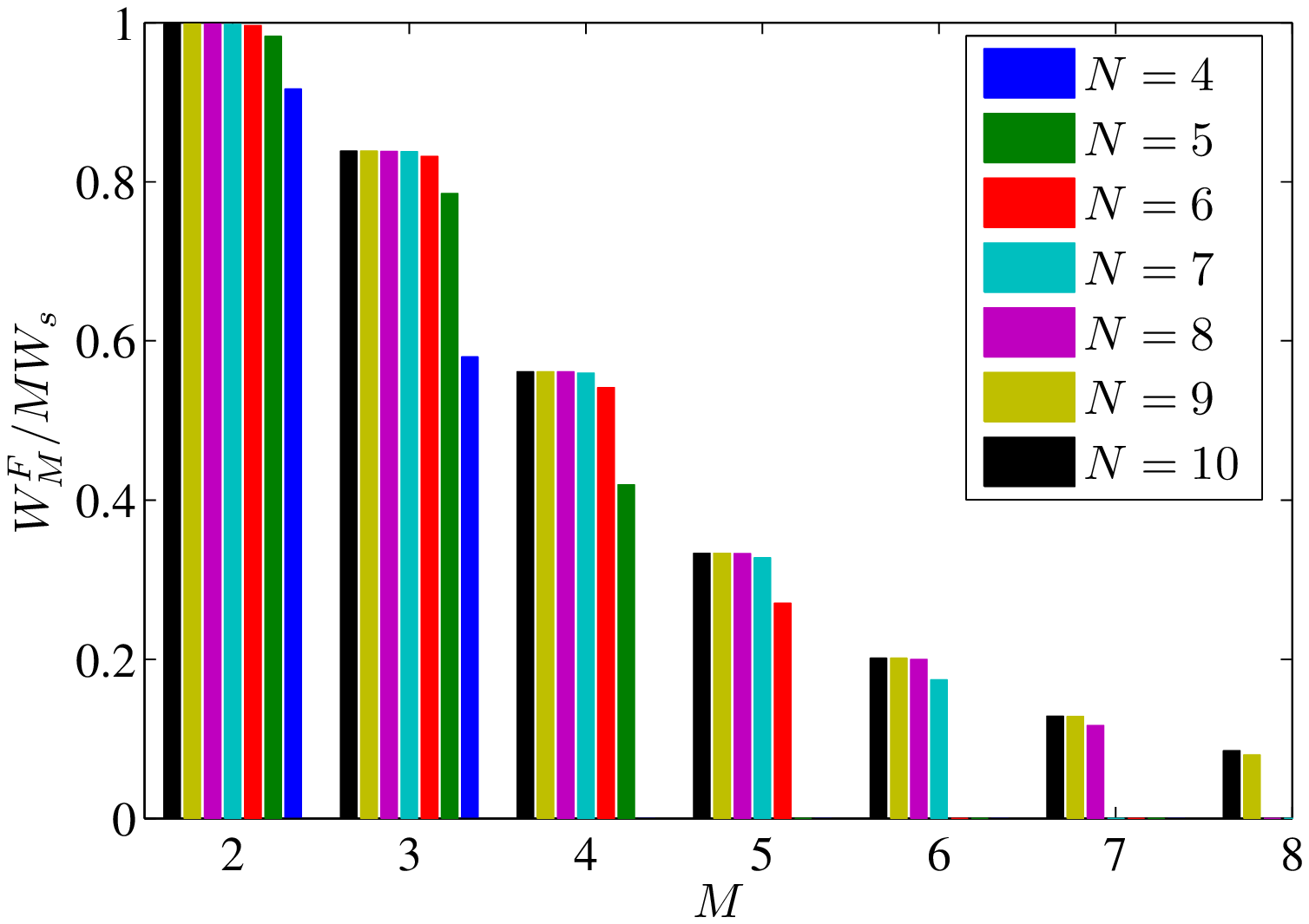}
\caption{(Color online) The ratios of $W^i_M/M$ to $W_s$ as a function of particle number $M$ for different energy level numbers $N$ in intermediate temperature regime. Other parameters are ${\hbar^2\pi^2}/{2mL_1^2kT_c}=1$, $R=2$ and $kT_h=5kT_c$ }\label{fig:70}
\end{figure}
We have studied the case of two identical particles as working substances. In this section, we will discuss the case of multiparticle. The crucial point of this discussion is the partition function of the canonical ensemble of $M$ particles. This can be done by following the result of Ref.\cite{Schmidt1998PA} as
\begin{eqnarray}
\mathcal Z_M^i(\beta)=\frac1M\sum_{m=1}^M(\pm1)^{m+1}\mathcal Z^i_1(m\beta)\mathcal Z^i_{M-m}(\beta),
\end{eqnarray}
where $\mathcal Z^i_0(\beta)=1$ and $i=B$ and $F$ respectively. Here the upper sign in the sum stands for Bosons ($i=B$) and the lower sign for Fermions ($i=F$) respectively. Based on this result, we can calculate the work done by the Otto heat engine with $N$-level $M$ identical particles as working substance. Here we also analyse the results in three regimes of different temperatures. In the low temperature regime, the results are easy, i.e., $W^B_M/W_s=1$ and $W^F_M/W_s=0$, where $W^B_M$ and $W^F_M$ are the work done by $M$ Bosons and Fermions respectively. This result does not depend on the energy level number $N$ and the particle number $M$ and it has a similar explanation to the two-particle case. The results of high temperature case are shown in Fig. \ref{fig:60}, where we plot $W^i_M/(MW_s)$ as a function of the particles number $M$ with different energy level $N$. $W^i_M/M$ can be seen as the average work done by a single particle in multiparticle heat engine. We see that  $W^B_M/(MW_s)>1$ under our condition, which means that for Bosons the average work done by a single particle in multiparticle heat engine is larger than a single particle as working substance under appropriate conditions. This ratio increases with $M$ but decreases with $N$. For Fermions, $W^F_M/(MW_s)<1$ and this ratio decrease with $M$ but increases with $N$. When $N$ is large enough, $W^i_M/(MW_s)\rightarrow1$ which means quantum statistics approaches to classical statistics in infinite energy level and high temperature limitations. But it is not true for low numbers of energy levels. In the intermediate temperature regime $W^i_M/(MW_s)<1$ under the conditions (Fig. \ref{fig:70}). For Bosons, it decreases with both $M$ and $N$. For Fermions, the ratio decreases with $M$ but increase with $N$.

\section{Discussions and conclusions} \label{sec:Con}
Before concluding this paper, we focus on other common models in quantum mechanics as  working substances. The first is particles trapped in harmonic potential, whose single particle energy spectrum is $E_n=\frac{n\hbar^2}{mL^2}$,
where $n$ starts from 0 and the zero point energy $\frac{\hbar^{2}}{2mL^{2}}$ is omitted. For this type of working substance, most results are similar to the ones given above. There are only two differences. One is when $N=M$ for Fermions, i.e. $(M+1)$-level $M$ particles case. In this case, $W^F_M/W_s=1$ for all other parameters, which means that the work done by $M$ Fermions equals to a single particle in $(M+1)$-level case. This is because the energy level structure of the $M$ Fermions is the same as this type of single particle energy spectrum. The other case is when $N\rightarrow\infty$, the work done by  identical Bosons $W^B_M$ is same as the work done by  identical Fermions $W^F_M$, i.e., $W^B_M=W^F_M=W$, which means that there is no difference between Bosons and Fermions in this case. This result does not depend on other parameters and can be proved analytically in two-particle cases as follows. Firstly, for this type of energy spectrum, we can obtain the partition functions of two Bosons and two Fermions in the limitation $N\rightarrow\infty$ as
\begin{eqnarray}
\mathcal Z_H^B&{=}&
\frac{1}
{\left[1{-}\exp\left({-}{\frac{\hbar^2}{mL^2kT}}\right)\right]^2\left[1{+}\exp\left({-}{\frac{\hbar^2}{mL^2kT}}\right)\right]}, \nonumber\\
\mathcal Z_H^F &{=}&
\frac{\exp\left({-}{\frac{\hbar^2}{mL^2kT}}\right)}
{\left[1{-}\exp\left({-}{\frac{\hbar^2}{mL^2kT}}\right)\right]^2\left[1{+}\exp\left({-}{\frac{\hbar^2}{mL^2kT}}\right)\right]}. \nonumber
\end{eqnarray}
Putting these two partition functions into Eqs. (\ref{eq:20}) and (\ref{eq:40}), we can calculate the work done by two Bosons or two Fermions are
\begin{eqnarray}
W=\frac{\hbar^2}{2mL_1^2}\left[3\coth\frac{\hbar^2}{mL_1^2kT_h}+\text{csch}\frac{\hbar^2}{mL_1^2kT_h}\right.  \nonumber\\
\left.-3\coth\frac{\hbar^2}{mL_2^2kT_c}-
\text{csch}\frac{\hbar^2}{mL_2^2kT_c}
\right].
\end{eqnarray}
As a result, the work done by two Bosons and two Fermions are the same. Qualitatively, this result can also be explained by the similarity of the energy level structure between  Bosons and  Fermions under the limitation $N\rightarrow\infty$ for energy spectrum of this type. Secondly, we should note that, although $W^B_M=W^F_M=W$ in this case, $W/MW_s\neq1$, which means that although the work done by the two Bosons and two Fermions are equal, and they are not equal to the results of  Boltzmann particles.


Another two common models are the extremely relativistic particle trapped in a box potential \cite{Pathriabook} and the particle trapped quartic potential \cite{Wang2011EPJD}, whose energy spectrum is $E_n=\frac{\pi\hbar c}{L}n$ and $E_n=\frac{\hbar^2}{mL^{4/3}}n^2$ respectively. After the same procedure, we can obtain the results of these two types of particles as working substances. We find that the results of particles trapped in quartic potential are similar to the ones of particles trapped in 1D box potential. Other two working substances, i.e., the extremely relativistic particle and particles trapped in harmonic potential are similar. The key point which determines the properties is the index of $n$.

It is another interesting problem to consider the results of a system composed of both Bosons and Fermions as working substances. The simplest case is one Boson and one Fermion. In this case, the two particles are different and they can be distinguished. As a result $W/W_s$ is 2, i.e., the result is classical. But when we have many Bosons and Fermions in one system as working substance, the case is very complex and it would be discussed elsewhere.

In conclusion, we have established  quantum Otto heat engines with  multilevel identical particles trapped in 1D box potential as the working substance. The efficiencies of multiparticle heat engines are the same to the ones of single-particle heat engine. Then in two-particle case, we concentrate on the ratios of $W^B$ or $W^F$ to $W_s$, and study the effects of other parameters on the ratios.  In the multilevel case, we consider three regimes. We find that the ratios are energy level number sensitive in high temperature and intermediate temperature regimes, but is not in low temperature regime. We also discuss the results of multiparticles case.
The results obtained here can be generalized to the working substances whose energy spectrum can be arranged as $E_n\sim n^2$. For the working substance in the form of $E_n\sim n$, we have two different results for the case of ($M+1$)-level $M$ Fermions and in the limitation $N\rightarrow\infty$.

\ \ \\
We thank  Yu-Han Gou for help. This work is supported by NSF of China under Grant Nos. 61475033 and 11605024  and the Foundation of Department of Education of Liaoning Province (L201683664).


\end{document}